\documentclass[]{article}
\usepackage{lmodern}
\usepackage{amssymb,amsmath}
\usepackage{ifxetex,ifluatex}
\usepackage{fixltx2e} % provides \textsubscript
\ifnum 0\ifxetex 1\fi\ifluatex 1\fi=0 % if pdftex
  \usepackage[T1]{fontenc}
  \usepackage[utf8]{inputenc}
\else % if luatex or xelatex
  \ifxetex
    \usepackage{mathspec}
  \else
    \usepackage{fontspec}
  \fi
  \defaultfontfeatures{Ligatures=TeX,Scale=MatchLowercase}
\fi
\IfFileExists{upquote.sty}{\usepackage{upquote}}{}
\IfFileExists{microtype.sty}{%
\usepackage{microtype}
\UseMicrotypeSet[protrusion]{basicmath} % disable protrusion for tt fonts
}{}
\usepackage{hyperref}
\PassOptionsToPackage{usenames,dvipsnames}{color} % color is loaded by hyperref
\hypersetup{unicode=true,
            pdftitle={Scientific notations for the digital era},
            colorlinks=true,
            linkcolor=ForestGreen,
            citecolor=red,
            urlcolor=blue,
            breaklinks=true}
\usepackage{color}
\usepackage{fancyvrb}

\DefineVerbatimEnvironment{Highlighting}{Verbatim}{commandchars=\\\{\}}
\newenvironment{Shaded}{}{}
\newcommand{\KeywordTok}[1]{\textcolor[rgb]{0.00,0.44,0.13}{\textbf{{#1}}}}

\newcommand{\DecValTok}[1]{\textcolor[rgb]{0.25,0.63,0.44}{{#1}}}

\newcommand{\ControlFlowTok}[1]{\textcolor[rgb]{0.00,0.44,0.13}{\textbf{{#1}}}}
\newcommand{\OperatorTok}[1]{\textcolor[rgb]{0.40,0.40,0.40}{{#1}}}

\newcommand{\NormalTok}[1]{{#1}}
\IfFileExists{parskip.sty}{%
\usepackage{parskip}
}{% else
\setlength{\parindent}{0pt}
\setlength{\parskip}{6pt plus 2pt minus 1pt}
}
\providecommand{\tightlist}{%
  \setlength{\itemsep}{0pt}\setlength{\parskip}{0pt}}
\ifx\paragraph\undefined\else
\let\oldparagraph\paragraph
\renewcommand{\paragraph}[1]{\oldparagraph{#1}\mbox{}}
\fi
\ifx\subparagraph\undefined\else
\let\oldsubparagraph\subparagraph
\renewcommand{\subparagraph}[1]{\oldsubparagraph{#1}\mbox{}}
\fi

\title{Scientific notations for the digital era}
\author{Konrad Hinsen \\ {\small Centre de Biophysique Moléculaire (UPR4301 CNRS)} \\ {\small Rue Charles Sadron, 45071 Orléans Cédex 2, France} \\ {\small konrad.hinsen@cnrs.fr} \and  \\ {\small Synchrotron SOLEIL, Division Expériences} \\ {\small B.P. 48, 91192 Gif sur Yvette, France} \\ {\small }} 
\date{}

\begin{document}
\maketitle
\begin{abstract}
Computers have profoundly changed the way scientific research is done.
Whereas the importance of computers as research tools is evident to
everyone, the impact of the digital revolution on the representation of
scientific knowledge is not yet widely recognized. An ever increasing
part of today's scientific knowledge is expressed, published, and
archived exclusively in the form of software and electronic datasets. In
this essay, I compare these digital scientific notations to the the
traditional scientific notations that have been used for centuries,
showing how the digital notations optimized for computerized processing
are often an obstacle to scientific communication and to creative work
by human scientists. I analyze the causes and propose guidelines for the
design of more human-friendly digital scientific notations.
\end{abstract}

\emph{Note}: This article is
\href{http://www.sjscience.org/article?id=527}{also available} in the
\href{http://www.sjscience.org/}{Self-Journal of Science}, where it is
open for public discussion and review.

\section{Introduction}\label{introduction}

Today's computing culture is focused on results. Computers and software
are seen primarily as tools that get a job done. They are judged by the
utility of the results they produce, by the resources (mainly time and
energy) they consume, and by the effort required for their construction
and maintenance. In fact, we have the same utilitarian attitude towards
computers and software as towards other technical artifacts such as
refrigerators or airplanes.

In scientific research, however, the path that leads to a result is as
important as the result itself. Drawing conclusions from an experimental
measurement requires a good understanding of the experimental setup that
was used to obtain the measurement. A scientist interpreting them must
know the reliability and precision of the devices that were used, and be
familiar with potential artifacts that could lead to misinterpretations.
Likewise, the computational results obtained from scientific software
can only be interpreted with a good understanding of what exactly the
software does. Scientific software therefore has the same status in
science as experimental setups and theoretical models.

In scientific discourse and in particular in the evaluation of a
research publication, results are therefore scrutinized together with
the path that lead to them. We expect experimentalists to explain the
materials and methods they have used, and theoreticians to explain their
reasoning in sufficient detail that their peers can understand it. We
should thus treat computational science in the same way and require
scientific software to be published and scrutinized in peer review.
While publication of scientific software is slowly becoming common, peer
review of this software remains exceptional. Its necessity is well
recognized in principle, but the effort required for such a review is
prohibitive. This is the most visible symptom of the problem that is the
topic of this essay. More generally, this problem is that digital
scientific knowledge is today expressed using notations such as
programming languages, which are not suitable for communication between
human scientists.

In the following, I will present a detailed analysis of this problem,
and propose some general guidelines for improving the situation. The
main audience is computational scientists, who have practical experience
with doing science using computers but no formal training in computer
science or in scientific epistemology. Readers with a computer science
background may skim over much of the second part. Note that I will not
propose \emph{the} solution to the problem, nor even \emph{a} solution.
My goal is to convince computational scientists that there \emph{is} a
problem, and that it \emph{can} be solved. Finding solutions that work
well is likely to require many years and the participation of many
people willing to test different ideas in practice.

The analysis that I present is applicable to all branches of science
whose models are based on continuous mathematics, such as algebraic,
differential, or integral equations. This includes almost all of physics
and chemistry, a good part of biology and the quantitative social
sciences, and all domains of applied research that build on foundations
in physics and chemistry. Much of what I say also applies to models
based on discrete mathematics, such as graphs or cellular automata, but
I will not consider them for the sake of simplicity. The examples I will
use for illustration reflect my own background in computational
biophysics, but readers shouldn't find it difficult to substitute
examples from their own field of work.

\subsection*{Outline}\label{outline}
\addcontentsline{toc}{subsection}{Outline}

I start by summarizing \protect\hyperlink{scientific-knowledge}{the
structure of scientific knowledge}, explaining factual, procedural, and
conceptual knowledge and why only the first two categories are used in
computation. Next, I outline how scientific communication and the
notations used for it \protect\hyperlink{evolution}{evolved in the
course of history}. These two sections prepare the discussion of
\protect\hyperlink{digital}{digital scientific knowledge} and why we
should care about it more than we do at the moment. This should provide
sufficient motivation for the reader to work through the more technical
\protect\hyperlink{formal-languages}{section on formal languages}, a
well-known concept in computer science that unifies two categories that
computational scientists tend to see as distinct: file formats for data
and programming languages, the two dominant forms of digital scientific
notation today.

After this more theoretical part, I explain the importance of
\protect\hyperlink{composition}{composition of information items} using
as an example the simulation of celestial mechanics, with an emphasis on
the constraints on the
\protect\hyperlink{composition-digital}{composition of digital
knowledge}. My goal is to illustrate what one should be able to do with
a proper digital scientific notation. I then compare to the
\protect\hyperlink{state-of-the-art}{state of the art} in computational
science, pointing out how it is inadequate for the study of
\protect\hyperlink{complex-systems}{complex systems}. One obstacle to
improvement is a perceived dichotomy between
\protect\hyperlink{software-data}{software and data}, which has its
roots in computing technology but has no counterpart in the structure of
scientific knowledge.

An important point that is often overlooked is the status of formal
languages as the main \protect\hyperlink{HCI}{human-computer interface}
in computational science. Doing research is a different task from
developing software, and requires a different interface. In particular,
we should pay more attention to the difference between
\protect\hyperlink{HCI-semantics}{human and computational semantics} and
to the need for simplicity and
\protect\hyperlink{flexibility}{flexibility} of a notation suitable for
humans doing creative research. Moreover, a digital scientific notation
must permit precise \protect\hyperlink{sr-references}{references to the
scientific record}.

In the last part of this essay, I consider solution strategies for the
problems that I have identified. I show two examples of how formal
languages can be made \protect\hyperlink{simple-and-flexible}{simple and
flexible} while providing a straightforward mechanism for composition:
\protect\hyperlink{XML}{XML} with its namespace mechanism for
composition, and the \protect\hyperlink{lisp}{Lisp} family of
programming languages with its macro system for creating small embedded
formal languages. I conclude by proposing
\protect\hyperlink{design-guidelines}{design guidelines} for digital
scientific notations.

\hypertarget{scientific-knowledge}{\section{The structure of scientific
knowledge}\label{scientific-knowledge}}

For the following discussion of scientific notation, it is useful to
classify scientific knowledge into three categories: factual,
procedural, and conceptual knowledge. Factual knowledge consists of the
kind of information one can record in tables, diagrams, or databases:
the density of water, the names of the bones in the human body, the
resolution of an instrument, etc. Procedural knowledge is about doing
things, such as using a microscope or finding the integral of a
function. Conceptual knowledge consists of principles, classifications,
theories, and other means that people use to organize and reason about
facts and actions.

Factual and procedural knowledge relies on conceptual knowledge. A table
listing the density of water at different temperatures refers to the
concepts of density and temperature. Instructions for using a microscope
refer to concepts such as sample or focus. Interpreting factual or
procedural knowledge requires a prior knowledge of the underlying
conceptual knowledge.

Conceptual knowledge has a hierarchical structure, with the definition
of every concept referring to more fundamental concepts. This leads to
the question of where this recursive process ends, i.e.~what the most
fundamental concepts are. When considering human knowledge as a whole,
this is a non-trivial problem in epistemology. For a discussion of
scientific knowledge, and in particular for the present discussion of
scientific notation, it is sufficient to consider the concepts of
everyday life as a given base level.

Factual and procedural knowledge often refer to each other. The
statement ``The orbit of the Moon around the Earth is reproduced to
precision A by solving Newton's equations for the solar system using
numerical algorithm B and initial values C'' is factual knowledge, once
specific A, B, and C are provided. But algorithm B is procedural
knowledge, which in turn refers to some other factual knowledge, such as
the masses of the Sun and its planets.

A final missing piece is metadata. Every piece of factual and procedural
knowledge comes with information attached to it that describes its
provenance and known limits of validity. A table showing the density of
water at different temperatures should state how, when, under which
conditions, and by who the listed values were obtained. It should also
provide an estimate of the values' accuracy.

In summary, the structure of scientific knowledge can be described as a
web of factual and procedural knowledge items that refer to each other,
and which are expressed in terms of concepts from the universe of
conceptual knowledge. The latter consists of multiple layers, with
concepts from everyday life at the bottom. Every layer refers to
concepts from lower, more fundamental layers.

\hypertarget{evolution}{\section{The evolution of scientific
communication}\label{evolution}}

Most of scientific communication takes place through research articles,
which are narratives that propose new factual and procedural knowledge,
occasionally also new concepts, and try to convince the scientific
community of the pertinence of this information. Over time, as a given
subject area becomes better understood, the scientific community usually
reaches a consensus about which concepts are the most useful for
describing its phenomena. The knowledge from a large number of research
articles is then distilled into review articles, monographs, and other
reference works. The knowledge considered most fundamental ends up in
textbooks for transmission to the next generation of scientists. Each
unit of scientific communication is written for a specific audience, and
relies on a stack of conceptual layers that this audience is expected to
be familiar with.

Before the use of computers, scientific knowledge was mainly recorded on
paper, using three forms of notation: written language, images, and
tables. Written text combines plain language, domain-specific
vocabulary, and shorthand notation such as mathematical formulas. Images
include both drawings and observations captured in photographs,
radiographs, etc. Tables represent datasets, which are most often
numerical.

There is a close relation between the conceptual knowledge on which a
narrative relies and the notation that it employs. Domain-specific
vocabulary directly names relevant concepts. Shorthand notation replaces
frequently used words and lengthy sentences that involve these concepts.
For example, Newton's laws of motion are commonly written as

\[F = m \cdot a\]

whose full-length equivalent is ``The force acting on a point mass is
equal to the product of its mass and its acceleration.'' Force, mass,
and acceleration are concepts from mechanics and \(F\), \(m\), and \(a\)
are conventional shorthands for them. The symbols \(=\) and \(\cdot\)
are shorthands for the concepts of equality and product, both of which
come from more fundamental conceptual layers in mathematics.

The standardization of scientific notation is variable and usually
related to the stability of the concepts that it expresses.
Well-established conceptual layers come with a consensus notation,
whereas young conceptual layers can be expressed very differently by
different authors. Scientists ``play around'' with both the concepts and
the notations in rapidly evolving fields, before eventually settling on
a consensus that has proven to work well enough. Even the most basic
aspects of mathematical notation that we take for granted today were at
some time the subject of substantial tinkering (1). Moreover, even a
consensus notation is not completely rigid. Personal and disciplinary
tastes and preferences are one cause of variation. As an example, there
are several common notations for distinguishing vectors from scalars in
geometry. Another cause is the limited number of concise names and
labels. For example, the preference for one-letter names in mathematical
formulas, combined with the useful convention of each letter having only
one meaning in a given document, often imposes deviations from consensus
naming schemes.

This pattern of a high variability during innovation phases giving way
to consensus as a field or technology matures is ubiquitous in science
and engineering. The time scale of the consolidation process is often
decisive for reaching a satisfactory consensus. The lack of consensus in
mature technology is felt as a nuisance by its users. A good example is
the pointless diversity in chargers for mobile phones. On the other
hand, premature consensus creates badly adapted technology that is
difficult to get rid of. Computing technology is particularly affected
by this problem. In fact, most of the standardized technology in
computing -- file formats, programming languages, file systems, Internet
protocols, etc. -- is no longer adequate for today's requirements. The
reason is that the technical possibilities -- and, as a consequence,
user demands -- evolve too fast for an orderly consensus formation,
whose time scale is defined by human cognitive processes and social
interactions that, unlike technological progress, have not seen any
spectacular acceleration.

\hypertarget{digital}{\section{Digital scientific
knowledge}\label{digital}}

In the context of computing, factual knowledge is stored in
\emph{datasets}, whereas procedural knowledge takes the form of
\emph{algorithms}. Conceptual knowledge is not affected by the
transition from manual to mechanized computation. Like research articles
and reference tables, datasets and algorithms implicitly refer to
conceptual knowledge to give meaning to their contents. However, the
concepts are not explicitly represented in the computer, because they
are not required to perform a computation. Applying algorithms to
datasets is a purely mechanical operation that does not require any
knowledge or understanding of the underlying concepts. What \emph{does}
require an understanding of the concepts is the verification that a
given computation is scientifically meaningful.

It is of course \emph{possible} to store and process conceptual
knowledge using computers, e.g.~in the form of
\href{http://en.wikipedia.org/wiki/Ontology_\%28information_science\%29}{ontologies},
which represent conceptual knowledge as factual knowledge at a different
semantic level. Such approaches are finding their place in scientific
communication in the form of \emph{semantic publishing} (2), whose goal
is to make the scientific record machine-readable and thus accessible to
automated analysis. However, performing a computation and managing
information \emph{about} this computation are different and independent
operations, just like using a microscope is a different activity from
researching the history of microscopy. I will come back to the role of
digital scientific notations in semantic publishing
\protect\hyperlink{sr-references}{later}.

This dissociation of mechanical computation from the conceptual
knowledge base that defines its meaning has been recognized as a problem
in various domains of digital knowledge management, for example in
database design (3). The typical symptom is the existence of electronic
datasets that nobody can interpret any more, because the original
designers of the software and data formats did not document their work
sufficiently well for their colleagues and successors. A frequent
variant is people modifying software and data formats without updating
the documentation. Every computational scientist has probably
experienced the difficulties of dealing with datasets stored in
undocumented formats, and with software whose inner workings are not
described anywhere in an understandable form.

The most vicious manifestation of this problem relates to scientific
software. Even when software is developed respecting the best practices
of software engineering, it may nevertheless compute something else than
what its users think it computes. Documentation can help to some degree
by explaining the authors' intentions to the users, but nothing permits
to verify that the documentation is complete and accurate. The only way
to make sure that users understand what a piece of software computes is
making the software's source code comprehensible for human readers.
Today, most scientific software source code is unintelligible to its
users, and sometimes it even becomes unintelligible to its developers
over time.

Some of the problems we are observing in computational science today are
direct consequences of the fact that scientists have an insufficient
understanding of the software they use. In particular, it suffers from
rampant software errors (4,5){]} and the near-universal
non-reproducibility of computational results (6,7). The scientific
community has failed so far to fully appreciate the double role of
scientific software as tools for performing computations and as
repositories of scientific knowledge (8). It has uncritically adopted
notations for digital knowledge that are not adapted to human
communication. As a consequence, the all-important critical discourse
that makes scientific research self-correcting in the long run does not
adequately cover digital scientific knowledge.

\hypertarget{formal-languages}{\section{Formal
languages}\label{formal-languages}}

The defining characteristic of digital scientific knowledge is the use
of
\href{http://en.wikipedia.org/wiki/Template:Formal_languages_and_grammars}{\emph{formal
languages}}, rather the the informal languages of human communication.
The term ``formal language'' is commonly used in computer science, but
in computational science we usually speak of ``data formats'', ``file
formats'', and ``programming languages'', all of which are specific
kinds of formal languages. In this section, I will give a minimal
overview of the characteristics of formal languages, which is necessary
for understanding their implications for digital scientific knowledge.

At the hardware level of a digital computer, a computation is a
multi-step process that transforms an input bit sequence into an output
bit sequence. Information processing by computers thus requires all data
to be expressed as bit sequences. Dealing with bit sequences is,
however, very inconvenient for humans. We therefore use data
representations that are more suitable for human brains, but still
exactly convertible from and to the bit sequences that are stored in a
computer's memory. These representations are called formal languages.
The definition of a formal language specifies precisely how some piece
of information is encoded in sequences of bits. Many formal languages
use text characters instead of bits for another level of convenience.
Since the mapping from text characters to bit sequences is
straightforward (the currently dominant mapping is called Unicode (9)),
this makes little difference in practice.

The definition of a formal language consists of two parts, syntax and
semantics. Syntax defines which bit patterns or text strings are valid
data items in the language. Syntax rules can be verified by a program
called a parser. Semantics define the \emph{meaning} of syntactically
correct data items. With one important exception, semantics are mere
conventions for the interpretation of digital data. As I explained
above, meaning refers to conceptual knowledge that a computer neither
has nor needs, since all it does is process bit sequences. The exception
concerns formal languages for expressing programs, i.e.~the rules used
by the computer for transforming data. The semantics of a programming
language define how each operation transforms input data into output
data. Writing down such transformation rules obviously requires a
notation for the data that is being worked on. For that reason, a
programming language also defines the syntax and semantics of data
structures. In fact, a programming language can express all aspects of a
computation. We use separate languages for data (``file formats'') only
as a convenience for users and for improving the efficiency of our
computers.

There is a huge number of formal languages today, which can be organized
into a hierarchy of abstraction layers, such that languages at a higher
level can incorporate languages from lower levels. As a simple example,
a programming language such as Fortran incorporates formal languages
defining individual data elements - integers, floating-point numbers,
etc. At the lowest level of this hierarchy, close to the bit level at
which computing hardware operates, we have formal languages such as
Unicode (9) for text characters or the floating-point number formats of
IEEE standard 754 (10). One level up we find the memory layout of
Fortran arrays, the layout of UTF-8 encoded text files, and many other
basic data structures and file formats. Structured file formats such as
XML (11) or HDF5 (12) are defined on the next higher level, as they
incorporate basic data structures such as trees, arrays, or text
strings. Programming languages such as Fortran or C reside on that level
as well.

Defining the semantics of a programming language is not a
straightforward task. For non-programming formal languages, semantics
are mere conventions and therefore defined by a document written for
human readers. The same approach can be adopted for a programming
language, resulting for example in the C language standard (13). But the
semantics of programs also matter for their execution by a computer, and
therefore a ``computer-readable'' definition of the semantics is
required as well. It takes the form of either a program that translates
the programming language into processor instructions, called a compiler,
or a program that directly performs the actions of the programming
language, called an interpreter. We thus have the C language standard
defining the semantics of the C language for human readers, and a C
compiler defining the semantics for execution by the computer.
Unfortunately, there is no way to ensure or verify that the two
definitions are equivalent. A computer program cannot do it, because the
C language standard is not written in a formal language. A human
computing expert cannot do it reliably, because a C compiler is much too
complicated for verification by inspection.

This is in fact the same situation as I described in the last section
for scientific software: the compiler is the equivalent of the
scientific software, and the language definition is the equivalent of
its documentation. This is not just a superficial analogy: there is in
fact no profound difference between a compiler and a piece of scientific
software. Both transform input data into output data according to
complex rules that are explained to human readers in a separate
document. Compilers are executable implementations of programming
languages in the same way as scientific software is an executable
implementation of scientific models. This analogy is useful because
computer scientists have invested considerable effort into bridging the
gap between executable and human-readable specifications of programming
languages. Most of the ideas and some of the tools developed in this
process can thus be adapted to scientific software.

The basic idea is to introduce
\href{https://en.wikipedia.org/wiki/Formal_specification}{\emph{formal
specifications}}, which are written in formal languages and thus
computer-readable, but which simpler than the software whose behavior
they specify, and therefore more comprehensible to human readers.
Specifications are simpler than the actual software for several reasons.
One of them is that a specification can neglect many usability issues of
software: performance, use of resources, portability between platforms,
user interfaces, etc. are all irrelevant for specifying the core
computations that the software performs. More simplification is possible
if one accepts mere \emph{definitions} instead of \emph{algorithms}. A
definition allows to test if a result is correct, but is not sufficient
to obtain a result. As a simple example, consider sorting. The
definition of a sorted list of items is ``an ordered list whose elements
are the same as those of the input list''. Any algorithm for actually
performing the sort operation is much more complicated. For a human
reader, the definition is usually sufficient to understand what is going
on, and testing procedures can verify that the algorithms implemented in
software actually conform to the definition.

Like specification languages, formal languages for representing digital
scientific knowledge must aim for simplicity to facilitate comprehension
by human scientists, in particular those not directly involved with the
development of scientific software. Much of the experience gained from
work on specification languages can probably be applied in the design of
formal languages for science, but there are also differences to be taken
into account. In particular, scientific knowledge differs from software
in that its principal purpose is not to compute something. Computation
in science is a means to an end which is understanding nature. In the
next section, I will show a few examples of scientific information items
and how they are used in the construction of scientific software while
also serving different purposes.

\hypertarget{composition}{\section{Composition of information
items}\label{composition}}

A key operation in information management is the composition of data
from various sources into a more complex assembly. Composition is a
well-known concept in software development, as software is usually
assembled from building blocks (procedures, classes, modules, \ldots{}),
including preexisting ones taken from libraries. But composition is also
an everyday task in the theoretical sciences, even though it is not
labeled as such and in fact rarely ever identified as a distinct
activity.

\subsection{Example: composing a model for the solar
system}\label{example-composing-a-model-for-the-solar-system}

Suppose you want to predict the positions of the planets of our solar
system over a few years. You would start with Newton's 17th-century
description of celestial mechanics and compose a model from the
following ingredients:

\begin{enumerate}
\def\labelenumi{\arabic{enumi}.}
\item
  Newton's law of motion: \(F = m \cdot a\)
\item
  Newton's law of gravitation:
  \(F_{ij} = G \frac{m_i \cdot m_j}{|r_i-r_j|^2}\)
\item
  The masses \(m_i\) of the sun and the planets.
\item
  A set of parameters, derived from past astronomical observations, to
  define the initial state.
\end{enumerate}

All these put together define the positions of the celestial bodies at
all times in the past and future. But each of these items has a meaning
independently of the others, and can be put to other uses, such as
computing how fast an apple falls to the ground. You can also use the
first two ingredients to prove energy conservation in celestial
mechanics, or to derive Kepler's laws. Moreover, each of these pieces
comes from a different source (observation, theoretical hypothesis,
\ldots{}) that requires a specific approach to validation. We want to be
able to compose them into a new entity called ``model for the solar
system'', but we also want each piece to retain its own identity for
other uses. Ideally, we want to present our solar system model as a
composition that references the individual ingredients. And in the
traditional printed-paper system of scientific communication, that's
exactly what we do.

Let's move on to computation. To make an actual prediction, you have to
add some more ingredients. The model as composed above only
\emph{defines} the planetary orbits, but doesn't tell you how to
\emph{compute} them. So you need to add:

\begin{enumerate}
\def\labelenumi{\arabic{enumi}.}
\setcounter{enumi}{4}
\item
  A numerical solver for ordinary differential equations (ODEs), such as
  Runge-Kutta.
\item
  Suitable parameters for that solver, depending on your accuracy and
  precision requirements. For Runge-Kutta, that's the size of the
  integration time step.
\item
  A finite-size number representation with associated rules for
  arithmetic, because you can't compute with real numbers.
\end{enumerate}

You can then take a large stock of pencils and paper and start to
compute. If you prefer to delegate the grunt work to a computer, you
need one final ingredient:

\begin{enumerate}
\def\labelenumi{\arabic{enumi}.}
\setcounter{enumi}{7}
\tightlist
\item
  A programming language, implemented in the form of a compiler or
  interpreter.
\end{enumerate}

Your final composition is then a simulation program for celestial
mechanics, made from eight distinct ingredients. Ideally, you would
publish each ingredient and the composition separately as nine
machine-readable nanopublications (14). Unfortunately, with the current
state of the art in computational science, that is not yet possible.

\hypertarget{composition-digital}{\subsection{Composition of digital
knowledge}\label{composition-digital}}

In the pre-digital era, composition was never much of a problem. A
scientist would take a few research articles or monographs describing
the various ingredients, and then write down their composition on a
fresh sheet of paper. Variations in the notations across different
sources would be no more than an inconvenience. Our pre-digital
scientist would translate notation into concepts when reading each
source, and the concepts into his or her preferred notation when writing
down the composition. As long as the concepts match, as they do in any
mature field of science, that is routine work.

Composition of digital knowledge is very different. The items to be
composed must be matched not only in terms of (human) concepts, but also
in terms of the syntax and semantics of a formal language. And that
means that all ingredients must be expressed in \emph{the same} formal
language, which is then also the language of the composed assembly.

If we start from ingredients expressed in different languages, we have
basically two options: translate everything to a common language, or
define a new formal language that is a superset of all the languages
used for expressing the various ingredients. We can of course choose a
mixture of these two extreme approaches. But both of them imply a lot of
overhead and add considerable complexity to the composed assembly.
Translation requires either tedious and error-prone manual labor, or
writing a program to do the job. Defining a superlanguage requires
implementing software tools for processing it.

As an illustration, consider a frequent situation in computational
science: a data processing program that reads a specific file format,
and a dataset stored in a different format. The translation option means
writing a file format converter. The superlanguage option means
extending the data processing program to read a second file format. In
both cases, the use of multiple formal languages adds complexity to the
composition that is unrelated to the real problem to be solved, which is
the data analysis. In software engineering, this is known as
``accidental complexity'', as opposed to the ``essential complexity''
inherent in the task (15).

As a second example, consider writing a program that is supposed to call
a procedure written in language A and another procedure written in
language B. The translation option means writing a compiler from A to B
or vice-versa. The superlanguage option means writing a compiler or
interpreter that accepts both languages A and B. A mixed approach could
use two compilers, one for A and one for B, that share a common target
language. The latter solution seems easy at first sight, because
compilers from A and B to processor instructions probably already exist.
However, the target language of a compiler is not ``processor
instructions'' but ``the processor instruction set plus specific
representations of data structures and conventions for code composition
and memory management''. It is unlikely that two unrelated compilers for
A and B have the same target language at this level of detail. Practice
has shown that combining code written in different programming languages
is always a source of trouble and errors, except when using tools that
were explicitly designed from the start for implementing the
superlanguage.

Many of the chores and frustrations in the daily life of a computational
scientist are manifestations of the composition problem for digital
knowledge. Some examples are

\begin{itemize}
\tightlist
\item
  file format conversion, as explained above
\item
  combining code in different languages, also explained above
\item
  software installation, which is the composition of an operating system
  with libraries and application-specific software into a functioning
  whole
\item
  package management, which is an attempt to facilitate software
  installation that re-creates the problem it tries to solve at another
  level
\item
  software maintenance, which is the continuous modification of source
  code to keep it composable with changing computational environments
\item
  I/O code in scientific software, which handles the composition of
  software and input data into a completely specified computation
\item
  workflow management, which is the composition of datasets with
  multiple independently written and installed software packages into a
  single computation
\end{itemize}

These examples should be sufficient to show that the management of
composition must be a high-priority consideration when designing formal
languages for digital scientific knowledge.

\hypertarget{state-of-the-art}{\section{The state of the art in managing
digital scientific knowledge}\label{state-of-the-art}}

In the last section I have listed the ingredients that need to be
combined in order to make a solar system simulator. Let's look at how
such a simulator is actually structured using today's scientific
computing technology. We have the following clearly identifiable pieces:

\begin{enumerate}
\def\labelenumi{\arabic{enumi}.}
\item
  A simulation program, written in a programming language such as
  Fortran or C++, which incorporates ingredients 1, 2, 5, 7, and 8.
\item
  An input file for that program, written in a special-purpose formal
  language defined by the author of the simulation program, containing
  ingredients 3, 4, and 6.
\end{enumerate}

The structure of the input file is usually simple, meaning that it is
straightforward to isolate ingredients 3, 4, and 6 from it. There is
even a good chance that the input file will permit annotation of these
items, indicating the sources they were taken from. If we are really
lucky, the formal language of the input file is documented and designed
to permit the extraction of information for other uses.

The simulation program itself is almost certainly a monolithic piece of
information that combines 1, 2, 5, 7, and 8 in an inextricable way. None
of the ingredients is easy to identify by inspection, and we'd better
not even envisage extracting them using computational tools for other
uses. If we want to change something, e.g.~use a different ODE solver or
a different finite-size number representation, we'd probably rewrite
large parts of the program from scratch. Worse, changing the finite-size
number representation might actually force us to rewrite the program in
a different language.

This is how today's scientific software is typically written, but let's
also look at what we \emph{could} do, using today's technology, if we
were making a special effort to maintain the modular structure of our
knowledge assembly.

The easiest part to factor out is number 5, the ODE solver. We could use
one from a program library, and even choose a library that proposes
several solvers. But using such a library comes at an additional cost in
combining all the parts. We have to write ingredients 1 and 2 according
to the rules of the library, and accept for 7 and 8 whatever the library
allows us to use. In fact, the library modifies the formal language we
use for writing our software, adding features but also imposing
constraints. Fortran plus ODEPACK is not the same language as Fortran on
its own.

Superficially, we can also factor out ingredients 1 and 2, which define
the equations fed to the ODE solver. We could isolate these ingredients
in the form of procedures (also called subroutines or functions). But
those procedures do \emph{not} represent the original equations. They
only represent one aspect of the equations: the numerical evaluation of
some of their subterms. We could not use these procedures to prove
energy conservation, nor to derive Kepler's laws.

Finally, we could envisage factoring out ingredient 7, the number
representation. For example, we could use a library such as MPFR (16) to
get access to wide range of floating-point formats. But the same remark
applies as for the use of an ODE library: we would have to translate
everything else into the C + MPFR language with its rather peculiar
requirements. Moreover, it's either MPFR or an ODE library, unless we
can find an ODE library written specifically for use with MPFR. The
reason why can't freely combine an ODE library with a finite-size
arithmetic library is the same that prevents us from using the
ODE-specific equation-evaluation procedures for other purposes: an ODE
library does not contain ODE solver algorithms, but specific
\emph{implementations} of such algorithms that are less versatile than
the algorithms themselves.

Leaving the narrow realm of development tools for numerical software, we
could try to factor out the equations, ingredients 1 and 2, using a
computer algebra system. Such a system lets us write down the equations
as such, not only the numerical computation of its subterms. While the
idea looks promising, the state of today's computer algebra systems
doesn't make this a practically useful approach. They are not designed
as parts of an ecosystem for scientific knowledge management. The formal
languages they use for expressing terms and equations are insufficiently
documented, and for commercial systems they are even partly secret. Some
computer algebra systems have export functions that generate numerical
code in a language like C or Fortran, but the exact semantics of this
export are again opaque for lack of documentation.

\hypertarget{complex-systems}{\subsection{Complex
systems}\label{complex-systems}}

For the example I have used in the last section, there is no real
problem in practice because the whole model is rather simple.
Ingredients 1 to 7 can be written down and composed on a single sheet of
paper. We use computers only because the actual computation is very long
to perform. It is quite feasible to do all theroretical work by hand,
and write a simulation program just for doing the computation. That was
in fact the dominant use of computers in science during their first few
decades.

The situation changes drastically when we consider complex systems. If
instead of the solar system we wish to simulate a protein at the atomic
scale, we use a model that is overall very similar except for the second
ingredient. Instead of Newton's law of gravitation, a one-line formula,
we have an expression for the interatomic forces made up of tens of
thousands of terms. The list of these terms is constructed from the
molecular structure by an algorithm, meaning that we need a computer --
and thus formal languages -- not only for \emph{simulating} our model
but already for \emph{defining} it. The model itself is digital
scientific knowledge.

Since we do not have adequate formal languages for writing down such
digital models today, we cannot express them at all. We cannot analyze
or discuss the model, nor compare it in depth to competing models. All
we have is research papers describing the design principles behind the
model, and software written to perform a numerical evaluation. The
software source code is impenetrable for anyone but its authors.
Moreover, there is obviously no way to verify that the software
evaluates the model correctly, because that would require some other
expression of the model for comparison. This is again an instance of the
problem that I discussed \protect\hyperlink{formal-languages}{earlier}
for the definition of the semantics of programming languages. Our model
should be part of the \emph{specification} of our software, rather than
being completely absorbed into its source code.

In the case of the popular models for biomolecular simulations, each of
them is implemented by several different software packages, with each
program producing somewhat different numbers for the same protein. On a
closer look, each program actually implements its own variation of the
model, with modifications made for performance reasons, or because the
software authors believe the modification to be an improvement. In the
end, what we think of as a model is really a family of different models
derived from common design principles. In the absence of human-readable
specifications of each variant, we cannot compile a detailed list of the
differences, let alone estimate their impact on the results we obtain.

Similar situations exist wherever scientific models have become too
complex to be written down on paper. As a second example, consider the
Community Earth System Model (17), a popular model for the evolution of
the Earth's climate. One would expect such a model to consist of a large
number of coupled partial differential equations describing the behavior
of the atmosphere and the oceans, and their complex interactions. But it
really is a software package that implements a numerical solver for the
equations. Contrary to the situation in biomolecular simulation, a
significant effort is made to ensure that this software package can be
considered a reliable reference implementation. But even if we trust the
software to reliably evaluate the model numerically, we have still lost
all the non-numerical uses of a scientific model.

\hypertarget{software-data}{\subsection{Software and data in
computational science}\label{software-data}}

It is customary in computational science to distinguish between computer
programs, also called \emph{software}, and the \emph{data} that these
programs process. But the above discussion of formal languages shows
that this distinction between software and data is not fundamental. We
could very well use a single language to define all aspects of a
computation, and obtain the result in the same language. This is in fact
very easy to do, by hard-coding all input data into the source code of
the program. In today's computing environments, that would be
inconvenient in practice, but that is mostly due to the way our tools
work.

From the point of view of digital knowledgement management, it is
desirable to identify the individual pieces of information we wish to
handle, and the operations we wish to perform on them. The above
analysis of a solar system simulation provides a simple example. We
would then design formal languages specifically as digital scientific
notations for our knowledge items. Software tools would be just tools,
consuming and transforming scientific knowledge but not absorbing it
into its source code. In other words, all scientific knowledge would
become \emph{data}.

Some recent developments can be seen as stepping stones towards this
goal. I will mention a single example, the specification of differential
equations in FEniCS (18). FEniCS is a software package that solves
partial differential equations numerically using the Finite Element
method. A feature that distinguishes FEniCS from similar software
packages is that it allows its users to write the differential equations
to be solved in a notation very similar to traditional mathematics. In
particular, the equations are written down as distinct information
items, i.e.~they are \emph{data}. They are \emph{not} absorbed into
program code that is structured according to the needs of software
development. Similar approaches are used in other mathematical software
packages. However, a crucial final step remains to be taken:
Differential equations for FEniCS are written in a FEniCS-specific
formal language that is not suitable for anything else than solving the
equations in FEniCS. The scientific knowledge must be reformulated to
fit to the tool. What we should have instead is a formal language for
expressing all aspects of differential equations, and many tools, FEniCS
being just one of them, that can process this formal language. In
particular, we would like to be able to \emph{compose} differential
equations describing some physical system from individual ingredients,
much like the equations governing the solar system are composed from the
law of motion and the law of gravity.

One psychological barrier to considering all scientific knowledge as
data is the fact that scientific knowledge includes algorithms. In the
example of the solar system simulation, the numerical method for solving
Newton's equation is an algorithm. The formal languages used to
represent data in computational science do not permit the expression of
algorithms. For most computational scientists, algorithms are parts of
programs, and thus expressed in a programming language. However, it is
easy to see that algorithms are just another kind of data. Compilers
translate algorithms from one formal language to another, i.e.~they
process algorithms as data. The same can be said of many tools we use
every day to develop and analyze software. The only novelty in my
proposal is that algorithms that count as scientific knowledge should be
available for all kinds of scrutiny \emph{in addition} to being
executable by a computer.

We can also envisage an intermediate stage in which software tools
continue to incorporate digital scientific knowledge just like they do
today, but in which we also express and publish all digital scientific
knowledge in human-friendly formal languages. The human-friendly version
would then be part of the software's specification, and the equivalence
of the two formulations would be verified as part of software testing.

\hypertarget{HCI}{\section{Human-computer interactions through formal
languages}\label{HCI}}

If computers are to be powerful tools for scientific research, the
computer's user interface must be designed to make the interaction of
scientists with computers fluent and error-free. Whereas most other uses
of computers happen through relatively simple interfaces (forms,
graphical representations, command lines, \ldots{}), the interface
between a scientist and a computer includes the formal languages in
which scientific information is encoded for computation. In this
respect, computational science resembles software development, where the
human-computer interface includes programming languages. This similarity
explains why techniques and tools from software engineering are
increasingly adopted by computational science.

It is widely recognized in software engineering that software source
code should be written primarily to explain the working of a program to
other programmers, with executability on a computer being a technical
constraint in this endeavor rather its main objective. Some authors even
go farther and claim that the human understanding of the program, shared
by the members of its development team, is the primary output of
software development, because it is what enables the team to maintain
and adapt the program as requirements evolve (19). Software engineering
research has therefore started to investigate the usability of
programming languages by programers (20).

In scientific research, human understanding takes an even more prominent
role because developing an understanding of nature is the ultimate goal
of science. Research tools, including software, are only a means to this
end. Digital scientific notations are the main human-computer interface
for research, and must be developed with that role in mind. The use of
formal languages is a technical constraint, but suitability for research
work and communication by human scientists must be the main design
criterion.

Today's digital scientific notations are programming languages and more
or less well-defined file formats. In this section, I will outline the
lessons learned from working with these adopted notations, and the
consequences we should draw for the design of proper scientific
notations in the future.

\hypertarget{HCI-semantics}{\subsection{Human vs.~computational
semantics}\label{HCI-semantics}}

A programming language fully defines the meaning of a program, and thus
completely defines the result of a computation.\footnote{At least it
  does in theory. The definitions of many popular languages are
  incomplete and ambiguous (21).} However, software source code has a
second semantic layer which matters only for human readers: references
to conceptual domain knowledge in the choice of identifiers. Mismatches
between what a program does and what its source code suggests it does
are a common source of mistakes in scientific software.

As an illustration, consider the following piece of Python code:

\begin{Shaded}
\begin{Highlighting}[]
\KeywordTok{def} \NormalTok{product(numbers):}
    \NormalTok{result }\OperatorTok{=} \DecValTok{1}
    \ControlFlowTok{for} \NormalTok{factor }\OperatorTok{in} \NormalTok{numbers:}
        \NormalTok{result }\OperatorTok{=} \NormalTok{result }\OperatorTok{+} \NormalTok{factor}
    \ControlFlowTok{return} \NormalTok{result}
\end{Highlighting}
\end{Shaded}

To a human reader, the names \texttt{product}, \texttt{numbers}, and
\texttt{factor} clearly suggest that this procedure multiplies a list of
numbers. A careful reader would notice the \texttt{+} sign, indicating
addition rather than multiplication. The careful reader would thus
conclude that this is a multiplication program containing a mistake.
This is exactly how a scientist reads formulas in a journal article:
their meaning is inferred from the meaning of all their constituents,
using an understanding of the context and an awareness of the
possibility of mistakes.

For a computer, the procedure shown above simply computes the sum of a
list of numbers. The identifiers carry no meaning at all; all that
matters is that different identifiers refer to different things. As a
consequence, the above procedure is executed without any error message.

If we analyze the situations that typically lead to program code like
the above example, the careful human reader turns out to be right: most
probably, the intention of the author is to perform multiplication, and
the plus sign is a mistake. It is highly unlikely that the author wanted
to perform an addition and chose multiplication-related terms to confuse
readers.

Since the program is perfectly coherent from a formal point of view,
approaches based on algorithmic source code analysis, such as type
checking, cannot find such mistakes. Software testing can be of help,
unless similar mistakes enter into both the application code and the
test code. Of today's software engineering techniques, the ones most
likely to be of help are pair programming and code review. Like peer
review of scientific articles, they rely on critical inspection by other
humans.

Code review is also similar to peer review in that it is reliable only
if the reviewer is an expert in the domain, even more so than the
original author. In the case of software, the reviewer must have a
perfect understanding of the programming languages and libraries used in
the project. This is not obvious from the above example, which is
particularly short and simple. Any careful reader will likely spot the
mistake, even without much programming experience. But more subtle
mistakes of this type do happen and do go unnoticed, in particular when
advanced language features are used that perhaps even the code's author
does not fully understand. As an example, few scientists with basic
Python knowledge are aware of the fact that the above five-line function
could in fact compute almost anything at all, depending on the context
in which it is used. All it takes is a class definition that defines
addition in an unexpected way.\footnote{Lest more experienced
  Pythonistas put up a smug grin reading this, I suggest they ask
  themselves if they fully understand how far the code's result can be
  manipulated from the outside using metaclasses. I am the first to
  admit that I don't.}

The main conclusion to draw from this is that digital scientific
knowledge must be written in terms of very simple formal languages, in
order to make human reviewing effective for finding mistakes. All the
semantic implications of a knowledge item must be clear from the
information itself and from the definition of the formal language it is
written in. Moreover, a scientist working in the same domain should be
able to read, understand, and memorize the language definition with
reasonable effort and ideally pick it up while acquiring the domain
knowledge itself, which is how we learn most of traditional scientific
notation.

\hypertarget{flexibility}{\subsection{Flexibility in scientific
notation}\label{flexibility}}

As I mentioned \protect\hyperlink{evolution}{above}, traditional
pre-digital scientific notation is the result of an evolutionary
process. In principle, scientists can use whatever notation they like,
on the condition that they explain it in their publication. However,
there is social pressure towards using well-established notation rather
than inventing new ones. In practice, this leads to variable notation
for new concepts that becomes more uniform over time as the concepts are
better understood and consensus is reached about representing them in
writing.

In contrast, formal languages used in computing are very rigid. The
reasons are numerous and include technical aspects (ease of design and
implementation) as well as historical ones (the advantages of
flexibility were not immediately recognized). Perhaps the biggest
barrier to flexibility left today is the near universal acceptance of
rigidity as normal and inevitable, in spite of the problems that result
from it. Most data formats used in computational science do not permit
any variation at all. When data formats turn out to be insufficient for
a new use case, the two possible choices are to ``bend the rules'' by
violating some part of the definition, or to define a new format. Since
bending the rules is often the solution of least effort in the short
run, many data formats become ambiguous over time, with different
software packages implementing different ``dialects'' of what everyone
pretends to be a common format.\footnote{Readers familiar with
  computational structural biology have probably had bad surprises of
  this kind with the PDB (22) format.} Since computer programs lack the
contextual background of humans, they cannot detect such variations,
leading to an erroneous interpretation of data.

Programming languages are vastly more complex than data formats. In
particular, implementing a programming language by writing of a compiler
or interpreter is a significant effort, and requires competences that
most computational scientists do not have. As a consequence, the
programming languages used for computational science are few in number.
Moreover, they are under the control of the individuals, teams, or
institutions that produce their implementations. For all practical
purposes, computational scientists consider programming languages as
imposed from outside. The only choice left to the individual scientist
or team is which of the existing languages to use, and then work around
its limitations.

A digital scientific notation should offer the same level of flexibility
as traditional scientific notation: a scientist should be able to state
``I use conventions X and Y with the following modifications'', defining
the modifications in a formal language to make them usable by computers.
Social pressure, e.g.~in peer review, would limit abuses of this
flexibility and lead to consensus formation in the long run.

\hypertarget{sr-references}{\subsection{References to the scientific
record}\label{sr-references}}

The main infrastructure of science as a social process is the
\emph{scientific record}, which consists of the totality of scientific
knowledge conserved in journal articles, monographs, textbooks, and
electronic databases of many kinds. Scientists refer to the scientific
record when they base new studies on prior work, but also when they
comment on work by their peers, or when they summarize the state of the
art in a review article, a monograph, or a textbook.

In scientific narratives, references to the scientific record are often
imprecise by citing only a journal article, leaving it to the reader to
find the relevant part of this article. It is, however, quite possible
to refer to a specific figure or equation by a number. For computational
work, references must be more precise: a dataset, a number, an equation.
A digital scientific notation must therefore encourage the use of small
information items that can be referenced individually while at the same
time keeping track of their context. It matters that composite
information items can be referenced as a whole but also permit access to
their individual ingredients, as I have illustrated in my
\protect\hyperlink{composition}{celestial mechanics example}.

The rapidly increasing volume of scientific data and facts is creating a
need for computer-aided analysis of the network of scientific knowledge.
This has motivated the development of \emph{semantic publishing} (2),
which consists in publishing scientific findings in a machine-readable
form where concepts become references to ontologies. Current research in
semantic publishing focuses on giving machine-readable semantics to
non-quantitative statements that are typically transmitted by the
narrative of a journal article. The development of digital scientific
notations that I wish to encourage by this essay can be seen as a
variant of semantic publishing applied to computational methods. In this
analogy, today's scientific software is similar to today's journal
articles in that neither form of expression permits the automated
extraction of embedded knowledge items.

\hypertarget{simple-and-flexible}{\section{Simple and flexible formal
languages}\label{simple-and-flexible}}

The criteria exposed in the \protect\hyperlink{HCI}{last section} lead
to a technical requirement for digital scientific notations: it must
accommodate a large number of small and simple formal languages and make
it straightforward to define variants of them. This may well seem
impossible to many computational scientists. Large, rigid,
general-purpose languages are today's standard for software development,
whereas small, rigid, and undocumented languages dominate scientific
data storage. However, there are examples of more flexible formal
languages, which can serve as a source of inspiration for the
development of digital scientific notations. I will describe two of them
in this section.

The main technical obstacle to flexibility in formal languages is the
requirement for composition that I have
\protect\hyperlink{composition-digital}{discussed earlier}: the
information items that enter into a composition must all be expressed in
the same language. If that condition is not satisfied, an additional
effort must be invested in the form of language conversion or more
complex software that can process multiple languages.

The solution is to design a \emph{framework} for a \emph{family} of
formal languages, and develop generic tools that can process any member
of this family and also compositions of different members. In other
words, flexibility enters the design and the support infrastructure at a
very early stage. This principle should become clearer from two concrete
examples: XML and Lisp.

\hypertarget{XML}{\subsection{XML: composable data formats}\label{XML}}

XML (11) is a framework for defining formal languages that express
tree-structured data. The central concept in XML is the \emph{element},
which is a node in a tree whose type is identified by a tag. The tag
also defines which attributes the element can have, and which conditions
its child elements must satisfy. A concrete XML-based data format is
defined by a \emph{schema}, which contains an exhaustive list of the
allowed tags and the constraints on the element types defined by each
tag. Given a data file and the schema it is supposed to respect, generic
XML processing tools can validate the data file, i.e.~check that it
conforms to the schema, and also perform many types of data
transformation that do not depend on the semantics of the data. Finally,
writing programs that do semantics-dependent processing is facilitated
by support libraries that take care of the semantics-independent
operations, in particular parsing and validating the incoming
information and producing correct result files. Because of these
advantages, XML has become very popular and a large variety of schemas
has been defined. Examples that may be familiar to computational
scientists are MathML and OpenMath for mathematical formulas, SVG for
vector graphics, CML for chemical data, and SBML for systems biology.

Composition of XML data means constructing a tree from elements defined
in different schemas. This was made possible with the introduction of
XML \emph{namespaces}. A single-schema XML document starts with a
reference to its schema. A multi-schema XML document lists multiple
schemas and associates a unique name with each of them. That name is
then prefixed to each tag in the document. This prefix ensures that even
in the presence of tag homonyms in the document's schemas, each element
has a unique and well-defined tag.

The XML namespace mechanism is an implementation of the superlanguage
approach that I \protect\hyperlink{composition-digital}{have described
earlier}. Processing such superlanguages is made straightforward because
the mechanisms for defining them are part of the XML definition. All
modern XML processing software implements namespaces, and therefore can
handle arbitrary superlanguages inside the XML universe.

XML namespaces are not a magical solution to composing unrelated data
items. Any software that performs semantics-dependent processing still
needs to deal with each schema individually. But the tasks of defining
languages, processing them, and processing compositions are enormously
simplified by the XML framework. Defining an XML schema is much simpler
than designing a complete data format, let alone a data format open for
extensions. Processing someone else's XML data is also much simpler than
processing someone else's ad-hoc data format, because the schema
provides a minimum of documentation. Finally, the namespace mechanism
encourages the definition of small schemas that can then be composed,
making well-designed XML-based data formats easier to understand for
human readers.

\hypertarget{lisp}{\subsection{Lisp: extensible programming
languages}\label{lisp}}

Most programming languages used today are constructed in much the same
way. A syntax definition specifies which sequences of text characters
are legal programs. This syntax definition is set in stone by the
language designer. Some syntactical elements define fundamental data
types, others fundamental executable operations. These basic building
blocks can be combined by the programmer into larger-scale building
blocks using language constructs for defining data structures,
procedures, classes, etc. In fact, programming is almost synonymous with
defining such entities and giving them names for later referring to
them. In other words, programming means extending the language by new
building blocks, the last of which is the program to be run. The
programmer cannot modify the syntax in any way, nor take any features
away from the basic language. This means in particular that the
programmer cannot make the language any \emph{simpler}.

One of the oldest family of programming languages, the Lisp family,
differs from this picture in an important way. Its syntax is defined in
two stages. The first stage merely defines how a central data structure
called a \emph{list} is written in terms of text characters. The
elements of a list can be any basic Lisp data type, e.g.~numbers or
symbols, but also other lists. Nested lists are equivalent to trees, and
in fact Lisp's nested lists are very similar to the trees of elements
that I have described in the \protect\hyperlink{XML}{section on XML}.
The second stage of Lisp's syntax defines which lists are legal
programs. The general convention is that the first element of a list
specifies a language construct to which the remaining elements are
parameters. For example, the list \texttt{(+\ 2\ 3)} means ``perform the
+ operation on the numbers 2 and 3'', whereas the list
\texttt{(define\ x\ (+\ 2\ 3))} means ``set variable x to the value of
the expression defined by the list \texttt{(+\ 2\ 3)}''.

This two-stage syntax is exploited in what is a very rare feature in
programming languages: the second syntax level can be modified by the
programmer, using a language construct called a \emph{macro}.
Technically, a macro is a function called as part of the compilation of
Lisp code. When the compiler hits a list whose first element specifies a
macro, it runs the macro function and substitutes the original
macro-calling list by the macro function's return value, which is then
compiled instead.

To understand the power of this construct, consider that a compiler is a
program that transforms another program written in language A into an
equivalent program written in language B. That is exactly what a macro
does: it translates a program written in some language M into basic
Lisp. The language M is defined by the macro itself, just like any
compiler is an operational definition of a language as I
\protect\hyperlink{formal-languages}{explained before}. Whatever the
macro accepts as arguments is a valid program in M. A macro thus
\emph{is} a compiler, and by defining macros a programmer can define his
or her own languages with no other restrictions than respecting the top
layer of Lisp's syntax, i.e.~the list syntax. Most macros merely define
small variations on the basic Lisp language, but nothing stops you from
writing a \texttt{fortran} macro to implement a language equivalent to
Fortran except that its syntax is defined in terms of nested lists.

The use of macros as building blocks of compilers has been pushed to a
very advanced level in the Racket language (23), a Lisp dialect which
its developers describe as a ``programmable programming language''. The
path from the first Lisp macros of the 1960s via Scheme's hygienic
macros to today's Racket has been a long one. For example, it turned out
that making macros composable is not trivial (24). Today's Racket
programming environment contains a large number of languages for various
purposes. Plain ``racket'' is a standard general-purpose programming
language. A core subset of ``racket'' is available as ``racket/base''.
Several languages are simplified forms of ``racket'' for teaching
purposes. The simplification does not merely take out language features,
but exploits the gain in simplicity for providing better error messages.
Other languages are extensions, such as ``typed/racket'' which adds
static type checking. But Racket also lifts the traditional Lisp
restriction of list-based syntax, providing a mechanism to write
language-specific parsers. Both Java and Python have been implemented in
Racket in this way. A language definition in Racket is nothing but a
library (25), meaning that any number of languages can co-exist.
Moreover, a new language can be based on any existing one, making it
straightforward to define small modifications.

A big advantage of the Lisp/Racket approach to implementing new
languages is that all those languages are interoperable, because they
are all compiled to basic Lisp/Racket. This is an implementation of the
translation approach to composing different languages that I have
\protect\hyperlink{composition-digital}{described before}. Another
advantage is that defining new languages becomes much easier.
Implementing a big language such as Python remains a difficult task even
in Racket. But implementing a small variation on an existing language --
take away some parts, add some others -- is simple enough to be
accessible to an average software developer.

\hypertarget{design-guidelines}{\section{Designing digital scientific
notations}\label{design-guidelines}}

The main conclusion from the analysis that I have presented in this
essay is that digital scientific notations should be based on formal
languages with the following properties:

\begin{itemize}
\item
  \textbf{Small and simple}: each formal language must be so small and
  simple that a scientist can memorize it easily and understand its
  semantics in detail.
\item
  \textbf{Flexible}: a scientist must be able to create modifications of
  existing languages used in his/her field in order to adapt them to new
  requirements and personal preferences.
\item
  \textbf{Interoperable}: composition of digital knowledge items
  expressed in different languages must be possible with reasonable
  effort.
\end{itemize}

The \protect\hyperlink{simple-and-flexible}{two examples} I have
presented above suggest that a good approach is to define a framework of
languages and implement generic tools for common manipulations. The
foundation of this framework should provide basic data types and data
structures:

\begin{itemize}
\tightlist
\item
  numbers (integers, rationals, floating-point, machine-level integers)
\item
  symbols
\item
  text
\item
  N-dimensional arrays
\item
  trees
\item
  sets
\item
  key-value maps (also called associative arrays, hash tables, or
  dictionaries)
\end{itemize}

The representation of these fundamental data types in terms of bit
sequences can be based on existing standards such as XML (text) or HDF5
(binary). It is probably inevitable to have multiple such
representations to take into account conflicting requirements of
different application domains. As long as automatic loss-less
interconversion can be ensured, this should not be an obstacle to
interoperability. An added advantage of keeping the lowest level of
representation flexible is the possibility to adapt to future
technological developments, for example IPFS (26) whose ``permanent
Web'' approach seems well adapted to preserving the scientific record.

There should also be a way to represent algorithms, but it is less
obvious how this should best be done. Any of the common Turing-complete
formalisms (lambda calculus, term rewriting, \ldots{}) could be used,
but it may turn out to be useful to have access to less powerful
formalisms as well, because they facilitate the automated analysis of
algorithms.

A next layer could introduce domain-specific but still widely used data
abstractions, e.g.~from geometry. For much of mathematics, the OpenMath
content dictionaries (27) could be adopted. On top of this layer, each
scientific community can build its own digital scientific notations, and
each scientist can fine-tune them to specific needs.

An illustration of how these principles can be applied is given by the
MOlecular SimulAtion Interchange Conventions (MOSAIC) (28), which define
a digital notation for molecular simulations. MOSAIC lacks the common
layer of data types listed above, and is therefore not easily
interoperable with other (future) digital notations. It does, however,
define data structures specific to molecular simulations in terms of
more generic data structures, in particular arrays. MOSAIC defines two
bit-level representations, based on XML and HDF5. A Python library (29)
proposes three further implementations in terms of Python data
structures, and implements I/O to and from the XML and HDF5
representations.

Traditional scientific notations have evolved as a byproduct of
scientific research, and digital scientific notations will have to
evolve in the same way in order to be well adapted to the task. In this
spirit, the ideas listed in this section are merely the basis I intend
to use in my own future work, but they may well turn out to be a dead
end in the long run. I would like to encourage computational scientists
to develop their own approaches if they think they can do better. As I
have stated in the introduction, my goal with this essay is not to
propose solutions, but to expose the problem. If computational
scientists start to think about ``digital scientific notation'' rather
than ``file formats'' and ``programming languages'', I consider my goal
achieved.

\section*{References}\label{references}
\addcontentsline{toc}{section}{References}

\hypertarget{refs}{}
\hypertarget{ref-ux5fhistoryux5f2016}{}
1. History of mathematical notation. In: Wikipedia, the free
encyclopedia {[}Internet{]}. 2016 {[}cited 2016 Apr 25{]}. Available
from:
\url{https://en.wikipedia.org/w/index.php?title=History_of_mathematical_notation\&oldid=714899577}

\hypertarget{ref-shottonux5fsemanticux5f2009}{}
2. Shotton D. Semantic publishing: The coming revolution in scientific
journal publishing. Learned Publishing {[}Internet{]}. 2009 Apr 1
{[}cited 2016 Apr 20{]};22(2):85--94. Available from:
\url{http://onlinelibrary.wiley.com/doi/10.1087/2009202/abstract}

\hypertarget{ref-borgidaux5fdataux5f2004}{}
3. Borgida A, Mylopoulos J. Data Semantics Revisited. In: Bussler C,
Tannen V, Fundulaki I, editors. Semantic Web and Databases
{[}Internet{]}. Springer Berlin Heidelberg; 2004 {[}cited 2016 Apr
19{]}. pp. 9--26. (Lecture notes in computer science). Available from:
\url{http://link.springer.com/chapter/10.1007/978-3-540-31839-2_2}

\hypertarget{ref-soergelux5frampantux5f2014}{}
4. Soergel DAW. Rampant software errors undermine scientific results.
F1000Research {[}Internet{]}. 2014; Available from:
\url{http://f1000research.com/articles/3-303/v1}

\hypertarget{ref-meraliux5fcomputationalux5f2010}{}
5. Merali Z. Computational science: .Error. Nature {[}Internet{]}.
2010;775--7. Available from: \url{http://dx.doi.org/10.1038/467775a}

\hypertarget{ref-stoddenux5fsettingux5f2013}{}
6. Stodden V, Bailey DH, Borwein J, LeVeque RJ, Rider W, Stein W.
Setting the Default to Reproducible {[}Internet{]}. 2013 Feb pp. 1--19.
Available from:
\url{http://icerm.brown.edu/tw12-5-rcem/icerm_report.pdf}

\hypertarget{ref-pengux5freproducibleux5f2011}{}
7. Peng RD. Reproducible research in computational science. Science
{[}Internet{]}. 2011;334(6060):1226--7. Available from:
\url{http://eutils.ncbi.nlm.nih.gov/entrez/eutils/elink.fcgi?dbfrom=pubmed\&id=22144613\&retmode=ref\&cmd=prlinks}

\hypertarget{ref-hinsenux5fcomputationalux5f2014}{}
8. Hinsen K. Computational science: Shifting the focus from tools to
models. F1000Research {[}Internet{]}. 2014;3. Available from:
\url{http://f1000research.com/articles/3-101/v1}

\hypertarget{ref-ux5funicodeux5f2015}{}
9. Unicode 8.0.0 {[}Internet{]}. 2015 {[}cited 2016 Apr 21{]}. Available
from: \url{http://www.unicode.org/versions/Unicode8.0.0/}

\hypertarget{ref-ux5fieeeux5f2008}{}
10. IEEE Standard for Floating-Point Arithmetic. IEEE Std 754-2008. 2008
Aug;1--70.

\hypertarget{ref-ux5fextensibleux5f1998}{}
11. Extensible Markup Language (XML) {[}Internet{]}. 1998--2016 {[}cited
2016 Apr 21{]}. Available from: \url{https://www.w3.org/XML/}

\hypertarget{ref-hdf5}{}
12. The HDF Group. Hierarchical Data Format, version 5 {[}Internet{]}.
1997--2016. Available from: \url{http://www.hdfgroup.org/HDF5/}

\hypertarget{ref-ux5fisoux2fiecux5f2011}{}
13. ISO/IEC 9899:2011 - Information technology -- Programming languages
-- C {[}Internet{]}. 2011 {[}cited 2016 Apr 21{]}. Available from:
\url{http://www.iso.org/iso/iso_catalogue/catalogue_tc/catalogue_detail.htm?csnumber=57853}

\hypertarget{ref-grothux5fanatomyux5f2010}{}
14. Groth P, Gibson A, Velterop J. The anatomy of a nanopublication.
Information Services \& Use {[}Internet{]}. 2010 Jan 1 {[}cited 2016 Apr
18{]};30(1-2):51--6. Available from:
\url{http://content.iospress.com/articles/information-services-and-use/isu613}

\hypertarget{ref-brooksux5fnoux5f1987}{}
15. Brooks FPJ. No Silver Bullet: Essence and Accidents of Software
Engineering. Computer. 1987 Apr;20(4):10--9.

\hypertarget{ref-fousseux5fmpfrux5f2007}{}
16. Fousse L, Hanrot G, Lefèvre V, Pélissier P, Zimmermann P. MPFR: A
Multiple-precision Binary Floating-point Library with Correct Rounding.
ACM Trans Math Softw {[}Internet{]}. 2007 Jun {[}cited 2016 Apr
21{]};33(2). Available from:
\url{http://doi.acm.org/10.1145/1236463.1236468}

\hypertarget{ref-ux5fcommunityux5f1983}{}
17. Community Earth System Model {[}Internet{]}. 1983--2016 {[}cited
2016 Apr 21{]}. Available from: \url{https://www2.cesm.ucar.edu/}

\hypertarget{ref-alnaesux5ffenicsux5f2015}{}
18. Alnæs M, Blechta J, Hake J, Johansson A, Kehlet B, Logg A, et al.
The FEniCS Project Version 1.5. Archive of Numerical Software
{[}Internet{]}. 2015 Dec 7 {[}cited 2016 Apr 21{]};3(100). Available
from:
\url{http://journals.ub.uni-heidelberg.de/index.php/ans/article/view/20553}

\hypertarget{ref-naurux5fprogrammingux5f1985}{}
19. Naur P. Programming as theory building. Microprocessing and
Microprogramming {[}Internet{]}. 1985 {[}cited 2016 Mar
31{]};15(5):253--61. Available from:
\url{http://www.sciencedirect.com/science/article/pii/0165607485900328}

\hypertarget{ref-ux5fplateauux5f2010}{}
20. PLATEAU '10: Evaluation and Usability of Programming Languages and
Tools {[}Internet{]}. New York, NY, USA: ACM; 2010. Available from:
\url{https://dl.acm.org/citation.cfm?id=1937117}

\hypertarget{ref-regehrux5fguideux5f2010}{}
21. Regehr J. A Guide to Undefined Behavior in C and C++ {[}Internet{]}.
2010 {[}cited 2016 Apr 20{]}. Available from:
\url{http://blog.regehr.org/archives/213}

\hypertarget{ref-wwpdbux5fatomicux5f2011}{}
22. wwPDB. Atomic Coordinate Entry Format Version 3.3 {[}Internet{]}.
2011 {[}cited 2016 Apr 21{]}. Available from:
\url{http://www.wwpdb.org/documentation/file-format-content/format33/v3.3.html}

\hypertarget{ref-plt-tr1}{}
23. Flatt M, PLT. Reference: Racket. PLT Design Inc. 2010. Report No.:
PLT-TR-2010-1.

\hypertarget{ref-flattux5fcomposableux5f2002}{}
24. Flatt M. Composable and Compilable Macros: You Want It when? In:
Proceedings of the Seventh ACM SIGPLAN International Conference on
Functional Programming {[}Internet{]}. New York, NY, USA: ACM; 2002
{[}cited 2016 Apr 14{]}. pp. 72--83. (ICFP '02). Available from:
\url{http://doi.acm.org/10.1145/581478.581486}

\hypertarget{ref-tobin-hochstadtux5flanguagesux5f2011}{}
25. Tobin-Hochstadt S, St-Amour V, Culpepper R, Flatt M, Felleisen M.
Languages As Libraries. In: Proceedings of the 32Nd ACM SIGPLAN
Conference on Programming Language Design and Implementation
{[}Internet{]}. New York, NY, USA: ACM; 2011 {[}cited 2016 Apr 6{]}. pp.
132--41. (PLDI '11). Available from:
\url{https://www.cs.utah.edu/plt/publications/pldi11-tscff.pdf}

\hypertarget{ref-benetux5fipfsux5f2014}{}
26. Benet J. IPFS - Content Addressed, Versioned, P2P File System. 2014
Jul 14 {[}cited 2016 Apr 26{]}; Available from:
\url{http://arxiv.org/abs/1407.3561}

\hypertarget{ref-openmathux5fsocietyux5fopenmathux5f2000}{}
27. OpenMath society. OpenMath {[}Internet{]}. 2000--2013 {[}cited 2016
Apr 21{]}. Available from: \url{http://www.openmath.org/index.html}

\hypertarget{ref-hinsenux5fmosaicux5f2014}{}
28. Hinsen K. MOSAIC: A data model and file formats for molecular
simulations. J Chem Inf Model {[}Internet{]}. 2014;54(1):131--7.
Available from:
\url{http://eutils.ncbi.nlm.nih.gov/entrez/eutils/elink.fcgi?dbfrom=pubmed\&id=24359023\&retmode=ref\&cmd=prlinks}

\hypertarget{ref-hinsenux5fkonradux5fpymosaicux5f2014}{}
29. Hinsen, Konrad. pyMosaic 0.3.1. 2014 {[}cited 2016 Apr 21{]};
Available from: \url{http://dx.doi.org/10.5281/zenodo.11648}

\end{document}